\documentclass[twocolumn,english,aps,article,showpacs]{revtex4}
\usepackage[latin9]{inputenc}
\usepackage{textcomp}
\usepackage{mathrsfs}
\usepackage{amsthm}
\usepackage{amsmath}
\usepackage{graphicx}
\usepackage{color}
\usepackage{stmaryrd}
\usepackage{soul}
\usepackage{amssymb}

\makeatletter

\providecommand{\tabularnewline}{\\}

\@ifundefined{textcolor}{}
{%
 \definecolor{BLACK}{gray}{0}
 \definecolor{WHITE}{gray}{1}
 \definecolor{RED}{rgb}{1,0,0}
 \definecolor{GREEN}{rgb}{0,1,0}
 \definecolor{BLUE}{rgb}{0,0,1}
 \definecolor{CYAN}{cmyk}{1,0,0,0}
 \definecolor{MAGENTA}{cmyk}{0,1,0,0}
 \definecolor{YELLOW}{cmyk}{0,0,1,0}
}
\numberwithin{equation}{section}
\numberwithin{figure}{section}

\usepackage[squaren]{SIunits}
\newcommand{\bra  }{\langle}
\newcommand{\ket  }{\rangle}
\newcommand{\up  }{\uparrow}
\newcommand{\down  }{\downarrow}
\newcommand{\TE}{\mathrm{TE}}
\newcommand{\TM}{\mathrm{TM}}

\makeatother

\usepackage{babel}
\begin{document}

\title{Origins and control of the polarization splitting in exciton-polaritons microwires}

\author{O. Lafont$^1$} 
\author{V. Ardizzone$^1$} 
\author{A. Lemaitre$^2$}
\author{I. Sagnes$^2$} 
\author{P. Senellart$^2$}
\author{J. Bloch$^2$}
\author{J. Tignon$^1$} 
\author{Ph. Roussignol$^1$} 
\author{E. Baudin$^1$} 
\affiliation{
$^1$Laboratoire Pierre Aigrain, \'Ecole Normale Sup\'{e}rieure - PSL Research University,
CNRS, Universit\'{e} Pierre et Marie Curie - Sorbonne Universit\'es, Universit\'{e} Paris Diderot - Sorbonne Paris Cit\'{e},
24, rue Lhomond, 75231 Paris Cedex 05, France\\
$^2$LPN/CNRS, Route de Nozay, F-91460 Marcoussis, France}

\date\today
\begin{abstract}
We report on the experimental investigation of the polarization-dependent
energy splitting in the lower exciton-polariton branches of a 1D microcavity.
The splitting observed for the lowest branch can reach up to \unit{1}{\milli\electronvolt}. It does not result from low temperature thermal constraints but from anisotropic mechanical internal strains induced
by etching. Those strains remove the degeneracy both in the photonic
({\normalsize{}$\delta E_{\mathrm{ph}}$)} and excitonic ({\normalsize{}$\delta E_{\mathrm{exc}}$)}
components of the polariton but also in the photon-exciton coupling
($\delta\Omega$). Those three contributions are accurately infered from experimental data. 
It appears that the sign and  magnitude of the
polarization splitting as well as the linear polarization of the corresponding 
polariton eigenstates can be tuned through the bare exciton-photon
detuning. Moreover, no dependence
on the width of the wire (from 3 to \unit{7}{\micro\meter}) is observed. We propose a mechanical
model explaining the universality of those observations paving the
way to the engineering of polarization eigenstates in microwires exciton-polaritons. 
\end{abstract}


\maketitle

\section{Introduction}
\label{sec-Introduction}

Semiconductor microcavities with embedded quantum wells (QWs) are able to confine both photons in the cavity formed by distribute	d Bragg reflectors (DBRs) and excitons in the quantum wells. In the strong coupling regime, when the exciton-photon dipolar coupling quantified by the Rabi coupling $\Omega$ exceeds the relaxation processes characteristic rate $\Gamma$, the fundamental excitations of the system are quasi-particles named microcavity polaritons. 
Those quasi-particles are mixed exciton-photon states of particular interest for
the study of degenerate quantum fluids (polariton condensates)~\cite{Carusotto2013} and non-linear quantum optics, in particular for the generation of twin photons.~\cite{Diederichs2007, Abbarchi2011}

By etching a planar microcavity in micrometer-sized wires, the polaritons are confined in a quasi-one dimensional (1D) structure. In the context of exciton-polariton condensates, this confinement can be modulated to tailor an effective 1D potential which can serve many applications like all-optical control of polariton flow~\cite{Wertz2010} or all-optical phase modulation in polaritonic circuits~\cite{Sturm2014}. 
Due to the lateral confinement of the electromagnetic field, several photonic modes couple to the exciton resulting in numerous exciton-polariton lower and upper branches. 
Moreover the resulting polariton eigenstates show an energy splitting between polarizations that are mainly parallel and orthogonal to the wire long axis, as observed by various authors.~\cite{Dasbach2002, Kuther1998, Tartakovskii1998, Ardizzone2012, Wertz2010}. This polarization-dependent splitting has proven to be a key-feature in different applications like optical parametric oscillation (OPO)~\cite{Ardizzone2012, Abbarchi2011}, or optical polariton interferometers~\cite{Sturm2014}. It plays a role in the edge states of polariton honeycomb lattices~\cite{Milicevic2015}, which were recently proposed as $\mathbb{Z}$-topological polariton insulators~\cite{Nalitov2015}. 
The possibility to tune this energy splitting is of particular interest for the development of microcavity polaritonic devices. 

While the polarization-dependent energy splitting is well-recognized and potentially useful, origins of this splitting and contributions of the different mechanisms identified by the various authors are still a subject of debate. Three mechanisms have been reported influencing the polaritons energies: 
Several authors have proposed that a polarization-dependent cavity energy is responsible for the polariton polarization-dependent energy splitting. The origin of this photonic splitting has been associated with an induced birefringence in the DBRs due to thermal strain~\cite{Diederichs2007} and with an anisotropic mode confinement~\cite{Tartakovskii1998}. 
A purely excitonic contribution has also been invoked~\cite{Dasbach2002}, its origins lying in the etched-induced strain release in the quantum well plane which leads to a reduction of the quantum well symmetry, and consequently lifts the  degeneracy  of the exciton manifold due to the exchange interaction. 
Finally, a polarization-dependent Rabi energy has also been observed on 2D microcavities by applying a stress with a tip~\cite{Balili2010}. The origins of this effect lies in the induced mixing of the heavy-hole exciton with the light-hole exciton, resulting in an anisotropic oscillator strength of the excitonic transition. 

\begin{figure*}[ht]
\includegraphics[width=17.5cm]{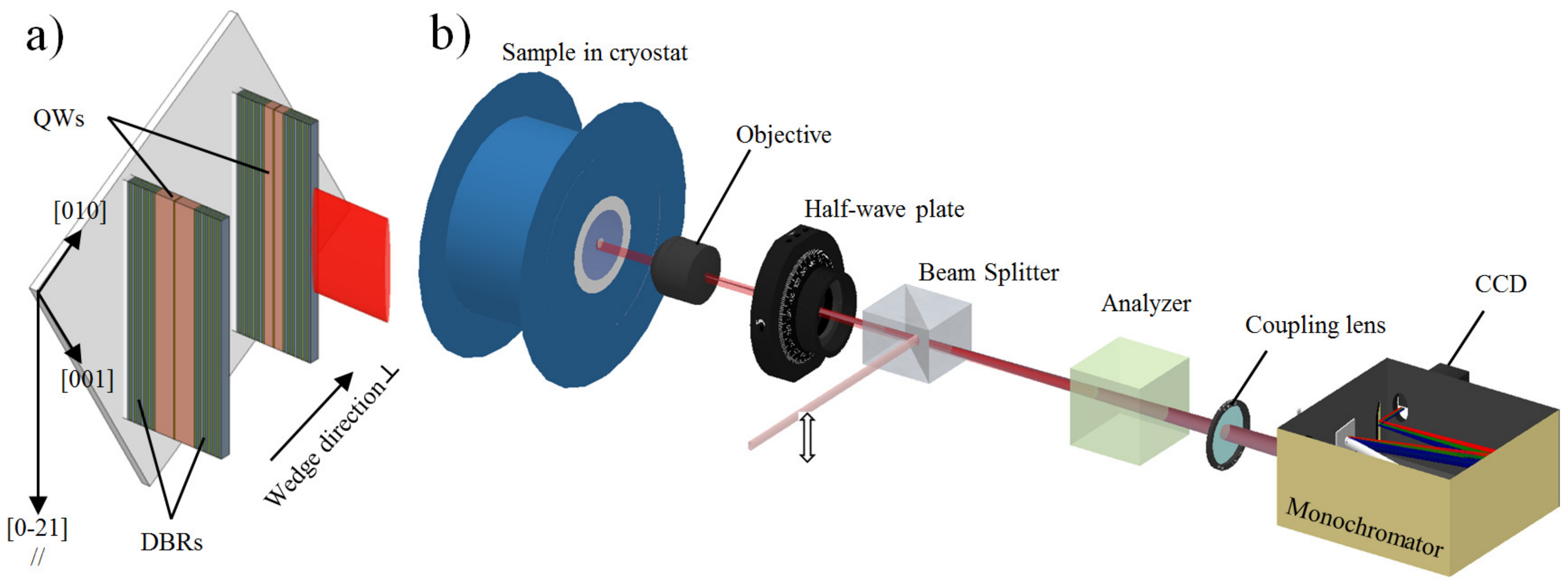}

\protect\caption{a) Schematic representation of the microwires microcavities. QWs in
the DBRs have been omitted by sake of simplicity. Wires are etched
down to the GaAs substrate approximately in the {[}210{]} direction
with respect to the cristallographic axis.\label{fig:Wire+Map} b)
Scheme of the polarization-resolved photoluminescence (PL) setup.
The sample is excited non resonantly with laser light blue-shifted
with regards to the polaritonic transitions in the first transmission
band of the DBR. Due to non polarization-free relaxation of the electronic
excitations, both TE (parallel) and TM (perpendicular) polarized polaritonic
branches are excited regardless of the polarization of the pump. By
using a combination of a half-wave plate and an analyzer, either the
TE-branch or the TM-branch is selected while always extinguishing the reflection
of the pump light. The laser spot is ellipsoidal in order to match as far as possible the wire geometry. }
\end{figure*}

In this work, experiments enlightening the contributions and origins of this polarization-dependent energy splitting in microwires are presented. The effects of the various external degrees of freedom are experimentally tested (bare exciton-photon detuning, temperature, applied stress, wire width, lateral mode confinement). 
Splittings up to \unit{1}{\milli\electronvolt} are observed, much larger than previously reported in the literature. 
Moreover, experiments reveal a change in the polarization splitting sign depending on the cavity-exciton detuning $\Delta$. 
These two observations - large magnitude and change of sign - suggest that the polarization splitting cannot be explained by a single contribution, either of photonic origin ($\delta E_{\mathrm{ph}}$), of excitonic origin ($\delta E_{\mathrm{exc}}$), or of the light-matter coupling ($\delta\Omega$). 
Our results show that all three splittings (photonic $\delta E_{\mathrm{ph}}$, excitonic $\delta E_{\mathrm{exc}}$, light-matter coupling $\delta\Omega$) contribute to the polarization splitting to a varying degree depending on the cavity-exciton detuning $\Delta$. Each contribution is carefully extracted from experimental data.

Beyond the relative importance of the different contributions for the fundamental polaritonic mode, these experiments point out their common origin through etched-induced strain relative to the microwire structure in Zinc-Blende crystal. 
This origin is further confimed by an analysis of the exact polarization basis for the polaritons eigenstates for various cavity-exciton detuning $\Delta$ along with a mechanical model of the microwires. 
An additional contribution due to lateral confinement is also evidenced for excited polaritonic modes. 
A precise model of the polarization splitting is developped and provides guidelines for the engineering of the microwires.

In section~\ref{sec-Experiment}, the experimental setup and the qualitative features of the polarization splitting determined experimentally are presented. In section~\ref{sec-Contributions}, the contributions of the photonic splitting, excitonic splitting, and light matter coupling splitting are quantified and discussed. Finally, in section~\ref{sec-Origins}, the inner strain of the microwires is proposed as a common origin for the polarization splitting and, as an experimental consequence, the control of the polarization axis of polariton branches is demonstrated. 
Confinement consequences are analyzed on excited lateral polaritonic modes. 
Possible ways to engineer the microwires are proposed.

\section{Experimental setup and qualitative features of the polarization splitting}
\label{sec-Experiment}

The sample used in this study is a microcavity formed by two $\lambda/2$ AlGaAs DBRs.
The DBRs consist in 26 (30) periodic alternance of $\textrm{Al\ensuremath{{}_{0.95}}Ga\ensuremath{{}_{0.05}}As}$
and $\textrm{Al\ensuremath{{}_{0.2}}Ga\ensuremath{{}_{0.8}}As}$ layers for the top (bottom) mirror.
Three groups of four \unit{7}{\nano\meter}-thick GaAs quantum wells are embedded in the cavity
and in the DBRs at the antinodes of the electric field to maximize
the light-matter coupling. During the MBE growth, the rotation of the
wafer is interrupted to introduce a wedge on the cavity thickness.
As a consequence, it is possible to tune the cavity modes energies
with respect to the excitonic mode energy by simply shifting the excitation
spot onto the sample. 
The wedge also affects the quantum well thickness, however, for a \unit{7}{\nano\meter}-thick GaAs quantum well, the dominant contribution to this energy variation is the exciton confinement energy of about $E_{\mathrm{conf}}=\unit{100}{\milli\electronvolt}$, the exciton binding energy being only a few meV. The confinement energy scales as $1/L_{\mathrm{QW}}^2$, where $L_{\mathrm{QW}}$ is the quantum well thickness and the cavity mode energy $E_{\mathrm{ph}}$ varies as $1/L_{\mathrm{cav}}$, where $L_{\mathrm{cav}}$ is the cavity effective thickness. Assuming a similar relative increase in the quantum well thickness and cavity effective length due to the wedge, we deduce that the exciton energy variation scales as $2 E_{\mathrm{conf}}/E_{\mathrm{ph}}$ where $E_\mathrm{ph}\simeq $\unit{1.6}{eV}.  This variation is thus only of the order of 13\% of the cavity energy variation and can be safely neglected.

The sample is etched in 1 mm-long parallel wires
of 3, 4, 5, 6 and \unit{7}{\micro\meter} widths with an approximate
depth of \unit{7}{\micro\meter}, \textit{i.e.}, down to the GaAs substrate. The wires long axis is orthogonal
to the direction of the wedge so that the cavity-exciton detuning
is the same all along the wire. 
The wires are etched in at \unit{39}{\degree} relative to the [100] GaAs crystalline axis. 

A single wire is non resonantly excited by using a Ti:Sapphire laser light
with an energy lying in the first nodes of the DBR reflection curve.
The excitation spot is elongated along the wire direction (approximate 
size:~\unit{50}{\micro\meter}$\times$\unit{5}{\micro\meter}) and the light beam impinges the sample at normal incidence. The order of magnitude of the excitation
power density is \unit{0.4}{\milli\watt \per \micro\meter \squared}. 
The photoluminescence (PL) from the sample is recorded using a CCD camera after a monochromator (see Fig.~\ref{fig:Wire+Map}-b). 
By using a collimating lens, the plan containing the entrance slit of the spectrometer is conjugated with the Fourier plane of the 
objective. This simple optical setup allows to observe the polariton dispersion curves. The linear polarization mainly parallel or orthogonal to the wire is selected with a half-wave plate and an analyzer. Note that during the experiment, the polarization in the detection path changes slightly with the detuning as the polarizer is not kept fixed but set to observe polaritonic branches in the polarization eigenmodes.

\begin{figure}[!h]
\includegraphics[width=8.6cm]{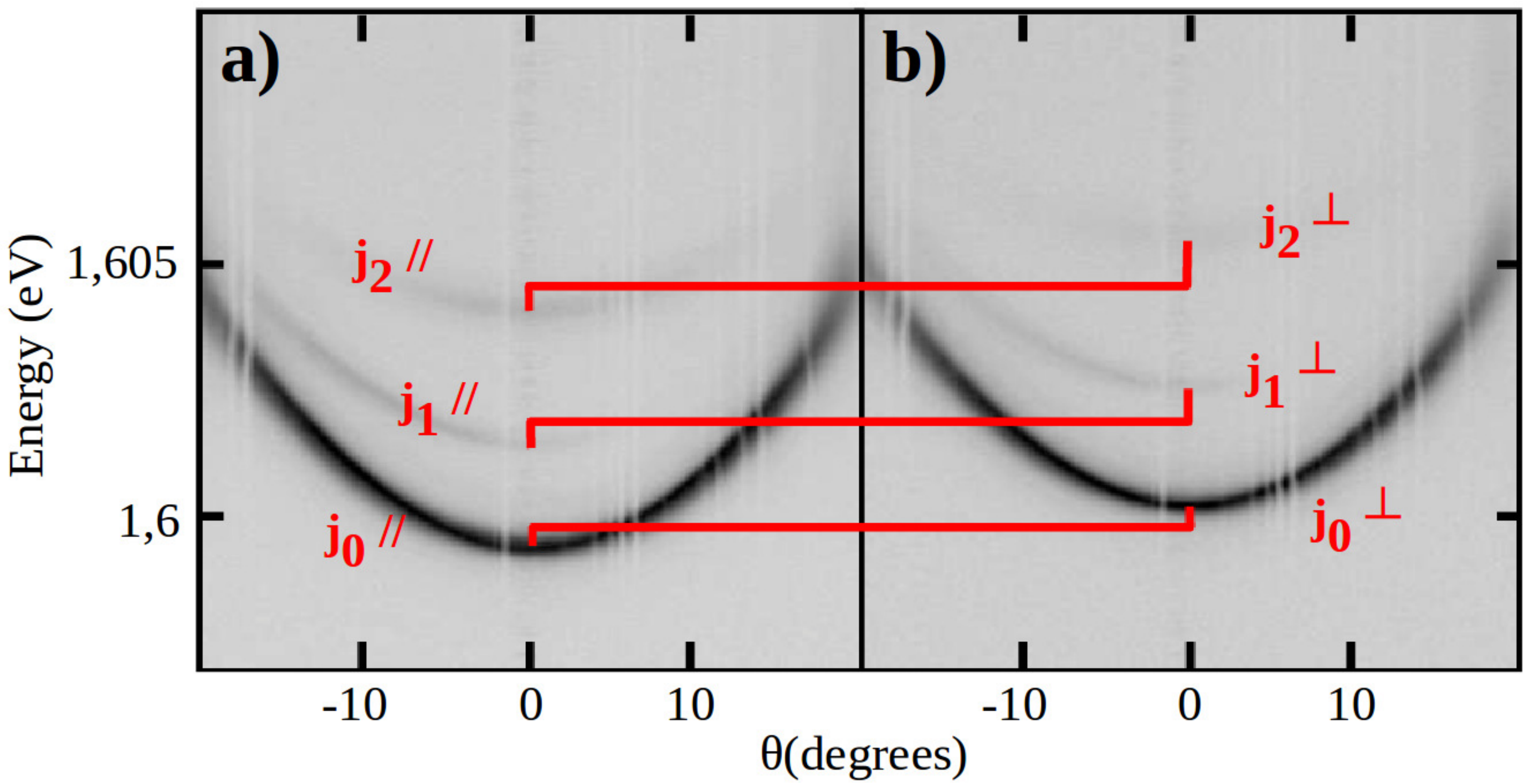}\protect\caption{Photoluminescence (PL) of the microstructure under non resonant excitation
observed in the reciprocal space. In this configuration, the dispersion
curves of the lower polarization branch of the 1D-confined microcavity
polaritons with polarization (a) parallel or (b) orthogonal  to the
wire axis can be directly observed. The splitting magnitudes for modes $j_0$, $j_1$ and $j_2$ are respectively \unit{820}{\micro\electronvolt}, \unit{1.17}{\milli\electronvolt} and \unit{1.33\milli\electronvolt.}.\label{fig:Splitting-1}}
\end{figure}

Figure~\ref{fig:Splitting-1} shows typical dispersion curves of the
lower polariton branches. An energy splitting between polarizations mainly
orthogonal and parallel to the wire long axis is visible. 
The polariton energies at $k=0$ are accurately obtained by fitting the corresponding polariton branches data with a model dispersion curve in order to take advantage of the redundancy of spectral information. The model used for the dispersion curve is a free fourth order polynomial which is sufficiently flexible to allow for resilience on the optical aberrations and possible misalignement (see appendix~E for details). 
The polarization-dependent energy splitting $\delta E_{\mathrm{pol}}$ is deduced by simply substracting the $k=0$ polariton energies in the polarization mainly orthogonal to the wire from the one mainly parallel to the wire, using the dispersion data previously obtained.

In the following, the experimental results are plotted as a function of $\Delta=E_{\mathrm{ph}}-E_{\mathrm{exc}}$ the cavity-exciton detuning. This reference quantity is independent on the wires width and therefore can be used to compare different data sets.
Since the regularly spaced microwires are etched out of a 2D microcavity possessing a linear cavity energy dependency on the position due to the wedge introduced during the growth step, the cavity-exciton detuning is therefore a linear function of the position on the sample. This linear relationship is confirmed by the linear asymptotic behaviour of the lowest polariton branch on Fig.~\ref{fig:Anticrossing}-a.

Let us now present the influence of the various degrees of freedom experimentally accessible on the polarization splitting. 

\subsection{Influence of external constraints}
\label{ssec-AppStress}

Stress applied to the sample can induce a birefringent behavior of the microcavity and possibly explain the appearance of the polarization splitting of polariton branches.~\cite{Diederichs2007} Besides, the finite wire width can be responsible for 1D excitonic/photonic confinement or strain release in the microwire. 

\begin{figure}[!h]
\includegraphics[width=8.7cm]{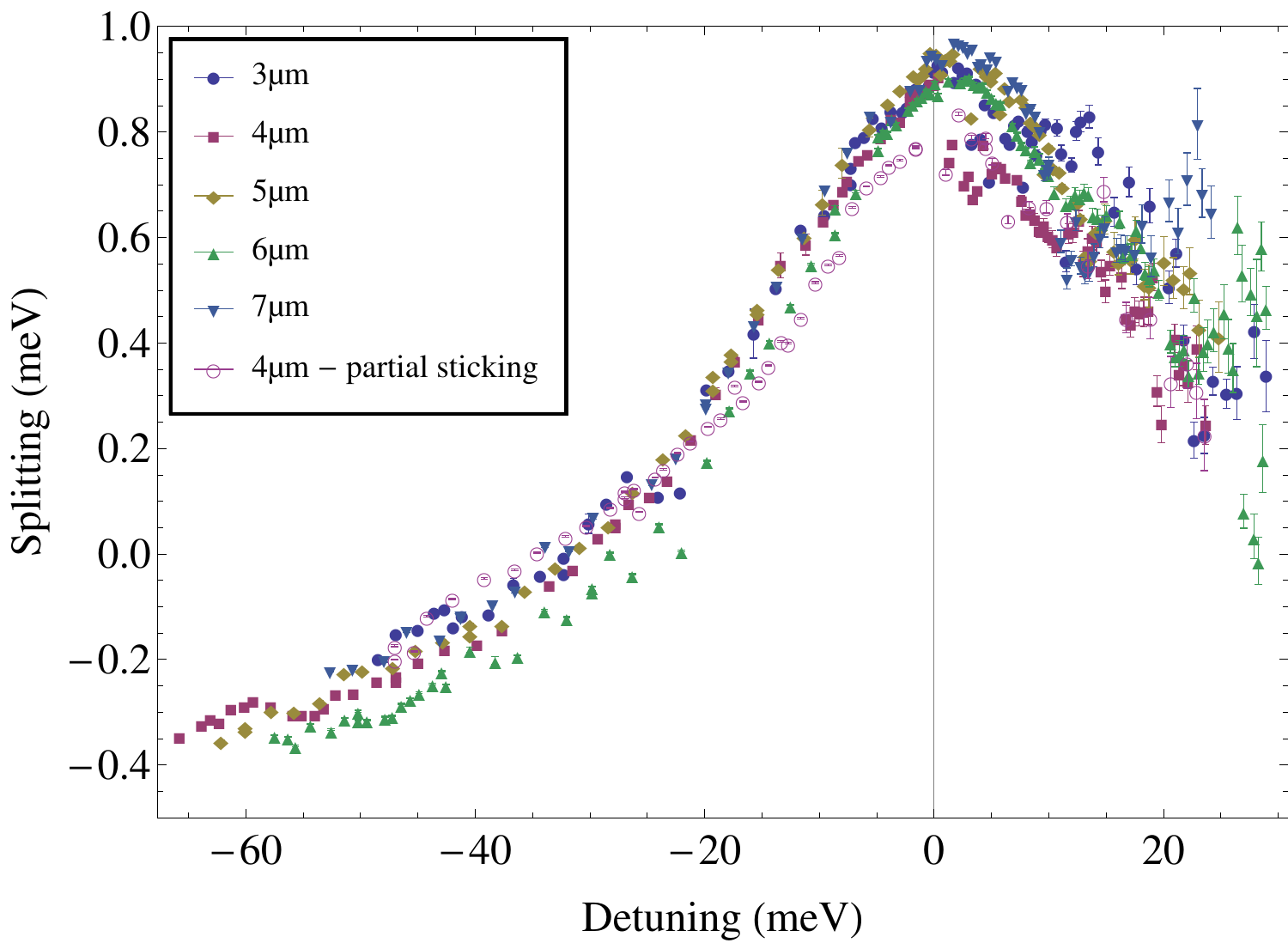}
\protect\caption{Polarization splitting as a function of the cavity-exciton detuning for various wire widths. Black dots corresponds to the measurement performed on \unit{4}{\micro\meter} wires when the sample is sticked far away from them to ensure that no thermal strain are induced. Each experimental dot corresponds to the measurement of one wire located at a different position on the sample, i.e. at different cavity-exciton detunings. \label{fig:Stick-1}}
\end{figure}

Figure~\ref{fig:Stick-1} represents the polarization splitting of the lowest polariton branch ($j_0$) measured on \unit{4}{\micro\meter}-wide wires when the sample is sticked on the cryostat cold finger via its whole surface or when it is sticked only on a small surface far away from the region of interest. This distant sticking (comparable to the size of the sample) guarantees that the thermally induced stress is reduced. No significant difference is observed between the two sticking configurations implying that strain in the sample due to thermal contraction of the sample holder does not contribute significantly to the polarization splitting observed. 

Figure~\ref{fig:Stick-1} shows also the polarization splitting as a function of detuning for various wire widths (3, 4, 5, 6 and \unit{7}{\micro\meter}). Again, no significant difference is observed between the various wire sizes.

These observations imply that the observed polarization splitting results from
local properties of the microcavities and not from boundary conditions
imposed by the wire width.

In appendix~E, a mechanical calculation of the structure response to applied stress shows that stress applied in the bulk material is relaxed in wires. In the range of widths explored in this work (below to \unit{7}{\micro\meter}) and knowing that the depth of etching is about \unit{7}{\micro\meter}, the mechanical model predicts that the lateral residual stresses are completely released in wires respectiveless of their widths. 

Let us now comment on the evolution of the observed polarization splitting with the detuning. 
At large negative detuning, the polariton state is mainly photonic and the negative observed splitting is a signature of the presence of a photonic contribution $\delta E_{\mathrm{ph}}$.
At null detuning, the Rabi energy plays an important role in the polariton total energy and the polarization splitting reaches a maximum positive peak. This suggests a contribution of the polarization Rabi splitting $\delta\Omega$. An excitonic one $\delta E_{\mathrm{exc}}$ can also be taken into account. These contributions are quantitatively analyzed in section~\ref{sec-Contributions}.

\subsection{Temperature dependence}
\label{ssec-Temp}

\begin{figure}[!h]
\includegraphics[width=8.6cm]{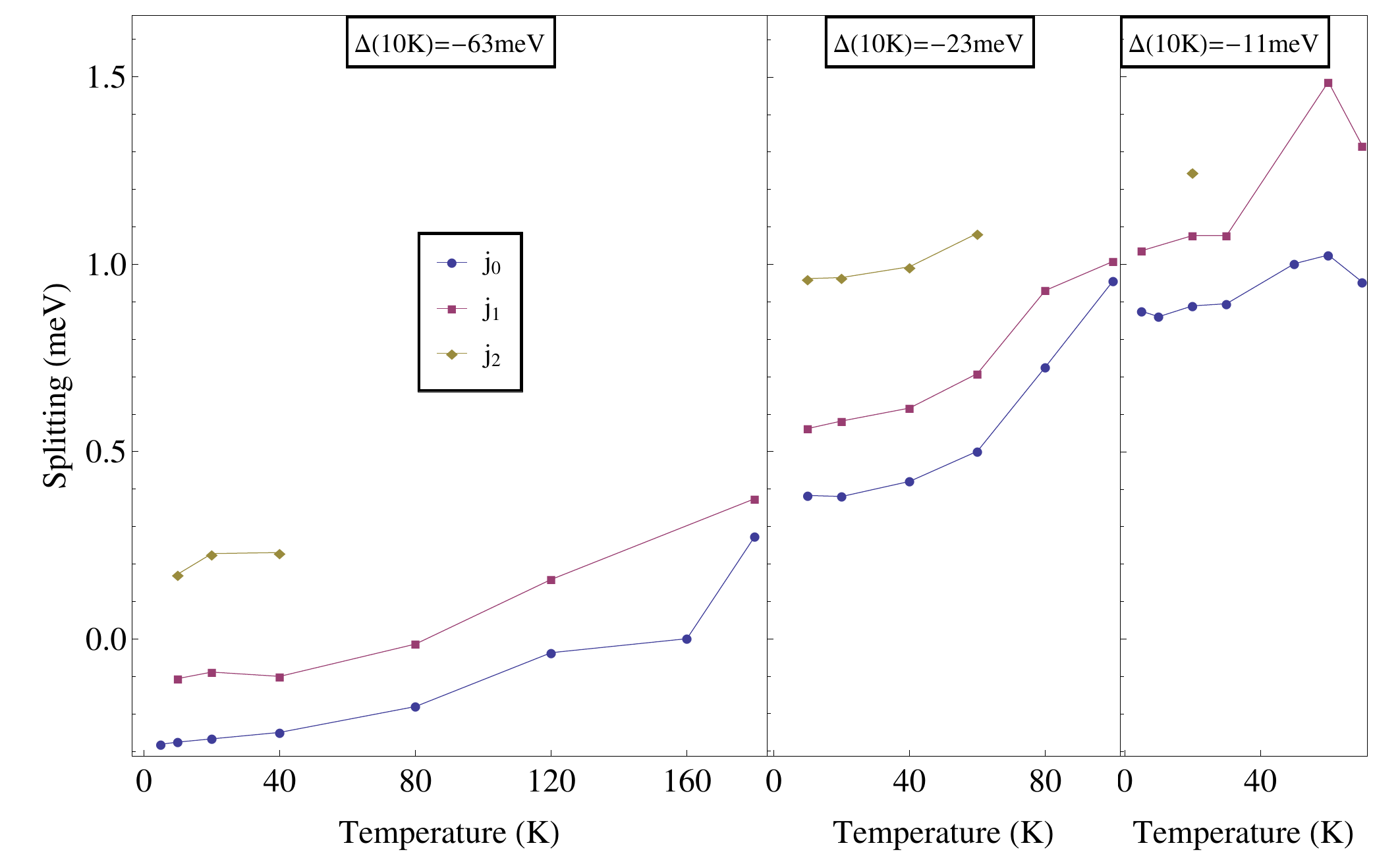}\protect\caption{Energy splitting as a function of the sample temperature for various detunings at \unit{10}{\kelvin} using \unit{3}{\micro\meter}-wide wires. Square, disk and triangle refer to the three lower polariton branches $j_{0}$, $j_{1}$ and $j_{2}$ respectively.\label{fig:Temperature-1}}
\end{figure}

Figure~\ref{fig:Temperature-1} displays the polarization splitting $\delta E_{\mathrm{pol}}$ as a function of the sample temperature for three bare detuning at \unit{10}{\kelvin} which are -63, -23 and \unit{-11}{\milli\electronvolt} respectively. 
These data are obtained considering the three first lower polariton branches $j_{0}$, $j_{1}$ and $j_{2}$ of three \unit{3}{\micro\meter}-wide wire. 

The polarization splitting increases with temperature, mode and detuning. This evolution can be simply understood in the light of the data depicted in Fig.~\ref{fig:Stick-1}: When the temperature increases, the GaAs bandgap is reduced and leads to a decrease in the exciton energy. If the cavity is negatively detuned from the exciton energy at low temperature, at high temperature the detuning gets closer to zero, in accordance with the observations of Fig.~\ref{fig:Stick-1}.
Besides, the different polaritonic modes $j$ correspond to the polaritons obtained from different photonic modes. Consequently, at negative detuning, an increase of $j$ is equivalent to an increase of the cavity energy and thus to a reduction of the cavity-exciton detuning. 
Taking into account the bandgap decrease with temperature, the values of the $j_0$ splitting with temperature is compatible with the ones of Fig.~\ref{fig:Stick-1}. 

\bigskip
The evolution of the polarization splitting with detuning is therefore highly reproducible and robust against external stresses.

\section{Contributions of the different source terms to the polarization splitting}
\label{sec-Contributions}

\begin{figure}[!h]
\includegraphics[width=8.6cm]{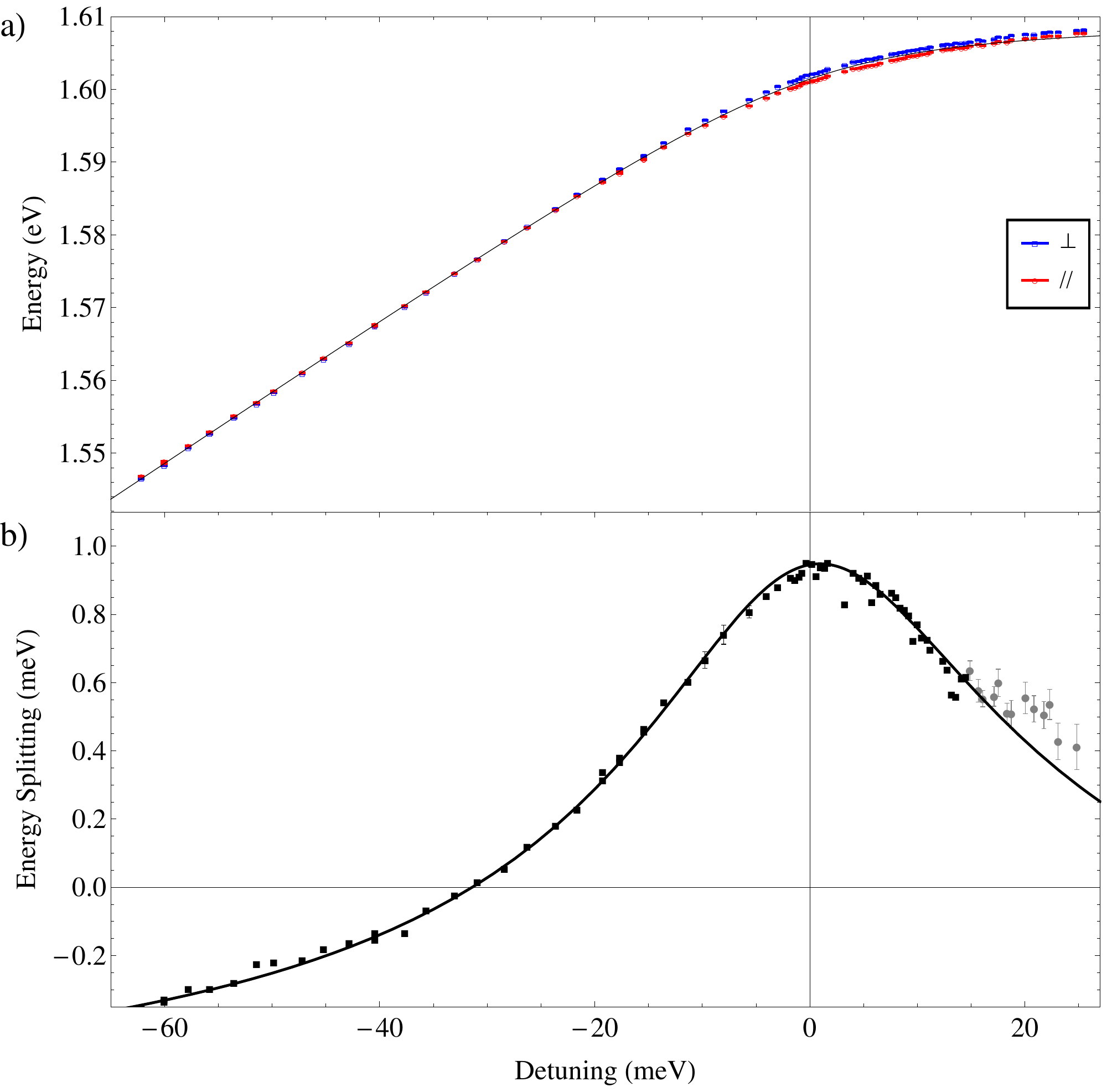}\protect\caption{a) Energy of the $k=0$ lower polariton for parallel (red open disks) and orthogonal
(blue open squares) polarization with respect to the wires axis for 82 \unit{5}{\micro\meter}-wires. Fit of the average energy as a function of detuning using eq.~\ref{eq:easy} is plotted as a plain line.\label{fig:Anticrossing} b) Polarization splitting mainly computed from Fig.~\ref{fig:Anticrossing}a. as a function of the cavity-exciton detuning. For large positive detunings (gray points), measurements are less accurate due to the broadening of the PL line: another analysis method described in appendix~B is used. The best non-linear fit using eq.~\ref{eq:splitting} is plotted as a plain line.}
\end{figure}

We first concentrate on the lowest polariton branch ($j_0$) which has the largest PL signal. Measurements have been performed on a set of 82 \unit{5}{\micro\meter}-wide wires. By carefully recording the
energy minima of the lower polariton dispersion curves on each wire, the anticrossing curve between the cavity mode and the excitonic mode can be obtained both for orthogonal and parallel polarization (Fig.~\ref{fig:Anticrossing}a.). 
From these two sets of measurements, the polarization energy splitting is precisely deduced as a function of the cavity-exciton detuning. From this dependency, the photonic, excitonic and Rabi contributions are extracted quantitatively. 

Figure~\ref{fig:Anticrossing}b. represents the energy splitting between parallely and orthogonally polarized branches. 
For negative and null detuning, these data are obtained by the energy difference of mainly parallely and orthogonally polarized polariton branches of Fig.~\ref{fig:Anticrossing}-a  when the deduced polarization splitting is sufficiently large compared to the polariton branches linewidth. For large positive detuning the polaritonic branch is mainly excitonic and its linewidth becomes large compared to the polarization splitting. In this case, the energy splitting is computed by another method described in appendix~B.

The anticrossings observed in Fig.~\ref{fig:Anticrossing}a can be accurately reproduced by the anticrossing relation (neglecting the influence of exciton and photon linewidth~\cite{Panzarini1999, Tassone1992}, this assumption being verified a posteriori, the photon and exciton linewidths being 0.1 and \unit{1}{\milli \electronvolt} respectively.): 
\begin{align}
E^{\substack{\perp \\ \sslash}} = & \frac 1 2 (E_{\mathrm{exc}}\pm \frac{\delta E_{\mathrm{exc}}}{2})+ 
\frac 1 2  (E_{\mathrm{ph}}\pm \frac{\delta E_{\mathrm{ph}}}{2})\nonumber\\
& - \frac 1 2 \displaystyle{\sqrt{(\Delta \pm \frac{\delta E_{\mathrm{ph}}}{2} \mp\frac{\delta E_\mathrm{exc}}{2})^2 + (\Omega\pm \delta\Omega)^2}}.
\label{eq-anticrossing}
\end{align}

However, considering that the Rabi energy is large compared to the polarization splitting terms, we rather fit a mean anticrossing energy averaged over the polarization:
\begin{equation}
E^{\mathrm{av}}=E_{\mathrm{exc}}+\frac \Delta 2 -\frac 1 2 \sqrt{\Delta^2+\Omega^2}
\label{eq:easy}
\end{equation}
Under the same assumption, the polarization splitting expression can be simplified, leading to the following expression:
\begin{align}
\delta E_{\mathrm{pol}}&=  E^{\perp} - E^{\sslash}\label{eq:splitting}\\
&\approx \frac 1 2 (\delta E_{\mathrm{exc}}+ \delta E_{\mathrm{ph}}) - \frac {A}{2} (\delta E_{\mathrm{ph}}- \delta E_{\mathrm{exc}}) -  B \delta \Omega,\nonumber
\end{align}
where $A=\frac {\Delta} {\sqrt{\Delta^2 + \Omega^2}}$ and $B=\frac {\Omega} {\sqrt{\Delta^2 + \Omega^2}}$. 

The mean polariton energy $E^{\mathrm{av}}$ is first fitted using~(\ref{eq:easy}) and the obtained value $\Omega$ is considered as a fixed parameter to fit the polarization splitting data with eq.~(\ref{eq:splitting}).
A multilinear fit of $\delta E_{\mathrm{pol}}$ as a function of $A$ and $B$ terms of eq.~(\ref{eq:splitting}) allows to infer the values of the polarization splitting contributions. 
This simple two-step fitting procedure allows to obtain realistic uncertainties on estimated parameters. A rigorous and complex single-step fitting procedure using the original polarization splitting expression~(\ref{eq-anticrossing}) has also been used leading to exactly the same results (see appendix~B).  

\begin{table}
\begin{tabular}{ccc}
Contribution & Value (meV) & Standard error (meV)\tabularnewline
\hline 
\hline 
$\delta E_{\mathrm{exc}}$ & -0.58 & 0.16 \tabularnewline
$\delta E_{\mathrm{ph}}$ & -0.76 & 0.07   \tabularnewline
$\delta\Omega$ & -1.62 & 0.12            \tabularnewline
\end{tabular}\protect\caption{Resulting contribution in meV from the different splitting sources obtained
by fitting the data for \unit{5}{\micro\meter}-wide wires with model (\ref{eq:splitting}) \label{tab:table} }
\end{table}

Table~\ref{tab:table} shows the contribution to the polarization splitting $\delta E_{\mathrm{pol}}$ infered from the data fitting procedure. 
These contributions have been independently identified  in previous works in various context: 
In ref.~\cite{Dasbach2005, Diederichs2007}, a splitting of $\delta E_{\mathrm{pol}} = -280$ $\mu$eV is reported in \unit{5}{\micro\meter}-wide wires. It is attributed to a small birefringence induced by thermal stress responsible for the energy splitting $\delta E_{\mathrm{ph}}$ of the photonic modes. 
In ref.~\cite{Dasbach2002}, a splitting of \unit{-130}{\micro\electronvolt} is reported in \unit{3}{\micro\meter}-wide wires and attributed to a small excitonic polarization splitting $\delta E_{\mathrm{exc}}$.  
The polarization splitting values reported in both studies~\cite{Dasbach2005,Dasbach2002} have been obtained while ignoring a possible Rabi polarization splitting. 
A posteriori, this is partly justified because (i) the samples used in both works were etched from the same microcavity which had a Rabi energy $\Omega=$\unit{3.8}{\milli\electronvolt}, whereas, in this work, a Rabi energy of \unit{16 \pm 1}{\milli\electronvolt} has been measured (ii) the wire etching in both works was done down to a layer located just below the QW, such that the strain release induced by etching was moderate due to the nearby influence of the bulk material (see Appendix~A). 
While the orders of magnitude reported in both papers are compatible with the ones reported in this work, a quantitative comparison is made difficult because of the critical dependence of the polarization splitting values with sample composition and etching depth, which are different from our sample (both samples used in the cited references were based on InGaAs ternary alloys whereas the sample studied in this work is made of AlGaAs ternary alloys). We shall develop this aspect in section \ref{sec-Origins} and in appendix~D.

The Rabi contribution has been identified in ref.~\cite{Balili2010} where an external stress is applied to a 2D microcavity by using a tip. This results in a heavy hole-light hole mixing of the excitonic states. The mixed excitonic states (polarized respectively perpendicular and parallel to the wire axis) have different oscillator strengths and give rise to the polarization Rabi splitting $\delta\Omega$. 
In ref~\cite{Balili2010}, the resulting splitting reached up to \unit{700}{\micro\electronvolt} but it can hardly be compared to the present study (\unit{1.6}{\milli\electronvolt}) due to the different origin of the strain. 

\section{Origins and control of the polarization splitting}
\label{sec-Origins}

In reference~\cite{Diederichs2007}, the polarization splitting was attributed to the thermal stress applied to the sample by the copper sample holder at low temperature. This origin was verified observing that the polarization splitting disappeared when the sample stood free from holder in an immersion cryostat. We have seen in section~\ref{ssec-AppStress} that this external stress was not at stake in the present sample.

However, a local strain must be responsible for this polarization splitting, such as internal strains due to the lattice mismatch between AlAs and GaAs layers. A mechanical model of the strain distribution in DBR layers, detailed in appendix~C, allows to describe accurately this effect. In particular, it shows that the lattice parameter along the wire is fixed by the lattice parameter of the wafer along the wire direction. As both the QWs and the wafer are made of GaAs, the strain vanishes in this direction, whereas, in the direction orthogonal to the wire, due to etching (see appendix~B), the average stress is released. Thus the lattice parameter of the semiconductor constituting the QW is fixed by the mean lattice parameters of the nearby DBRs. 

\subsection{Polarization splitting due to etching-induced strain}
\label{ssec-Origins-strain}

In appendix~D, the polarization excitonic, photonic and Rabi splitting from strain origin are deduced from first principles using the strain resulting from etching.

The deduced polarization splittings are given by $\delta E_{\mathrm{ph}}=\frac{E_{\mathrm{ph}} \delta \eta}{2 n^{(2)}}$, $\delta E_{\mathrm{exc}} \sim \frac{3}{2}  \frac{I_{hl} \Delta E_{\mathrm{SR}}}{|\phi^{\mathrm{3D}}(0)|^2} \frac{b \epsilon_{xx}^{(0)}}{\Delta E}$, and $\delta \Omega \sim \frac{\Omega b \epsilon_{xx}^{(0)}}{\Delta E}$, 
where $E_{\mathrm{ph}}$ is the bare cavity mode energy, 
$\delta \eta$ is the birefringent coefficient of the strained intracavity material defined in appendix~D, 
$n^{(2)}$ is the intracavity material refractive index, 
$\Delta E_{\mathrm{SR}}$ is the short range exchange splitting in the bulk semiconductor \cite{Maialle1993}, 
$\phi^{\mathrm{3D}}(0)$ is the 3D hydrogenic exciton wave function at zero relative distance,
$I_{hl}$ is a form factor defined in appendix~F,
$b$ is the deformation potential in the Pikus-Bir Hamiltonian, 
$\epsilon_{xx}^{(0)}$ is the strain of the QW material in the wire lateral direction, 
and $\Delta E$ is the light hole-heavy hole energy splitting. 
The corresponding polarization splittings obtained for the microwires considered in this work are $\delta E_{\mathrm{ph}}\sim\unit{-1.0}{\milli \electronvolt}$ (typical value), $\delta E_{\mathrm{exc}}\sim\unit{-0.38}{\milli \electronvolt}$, and $\delta \Omega \sim\unit{-1.0}{\milli \electronvolt}$ respectively (see appendix~F for theoretical values used). 
The three polarization splittings are correctly reproduced by this model. The ~35\% difference between experimental and theoretical values for $\delta \Omega$ and $\delta E_{\mathrm{exc}}$ can be attributed to the poor estimation of the light hole-heavy hole exciton splitting $\Delta E$ which is extremely sensitive to both composition and size of the quantum wells, whereas the overestimation of $\delta E_{\mathrm{ph}}$ is due to the use of extrapolated room temperature photoelastic coefficients for the DBR materials.\\

\begin{center}
\begin{figure}[h]
\includegraphics[width=8.6cm]{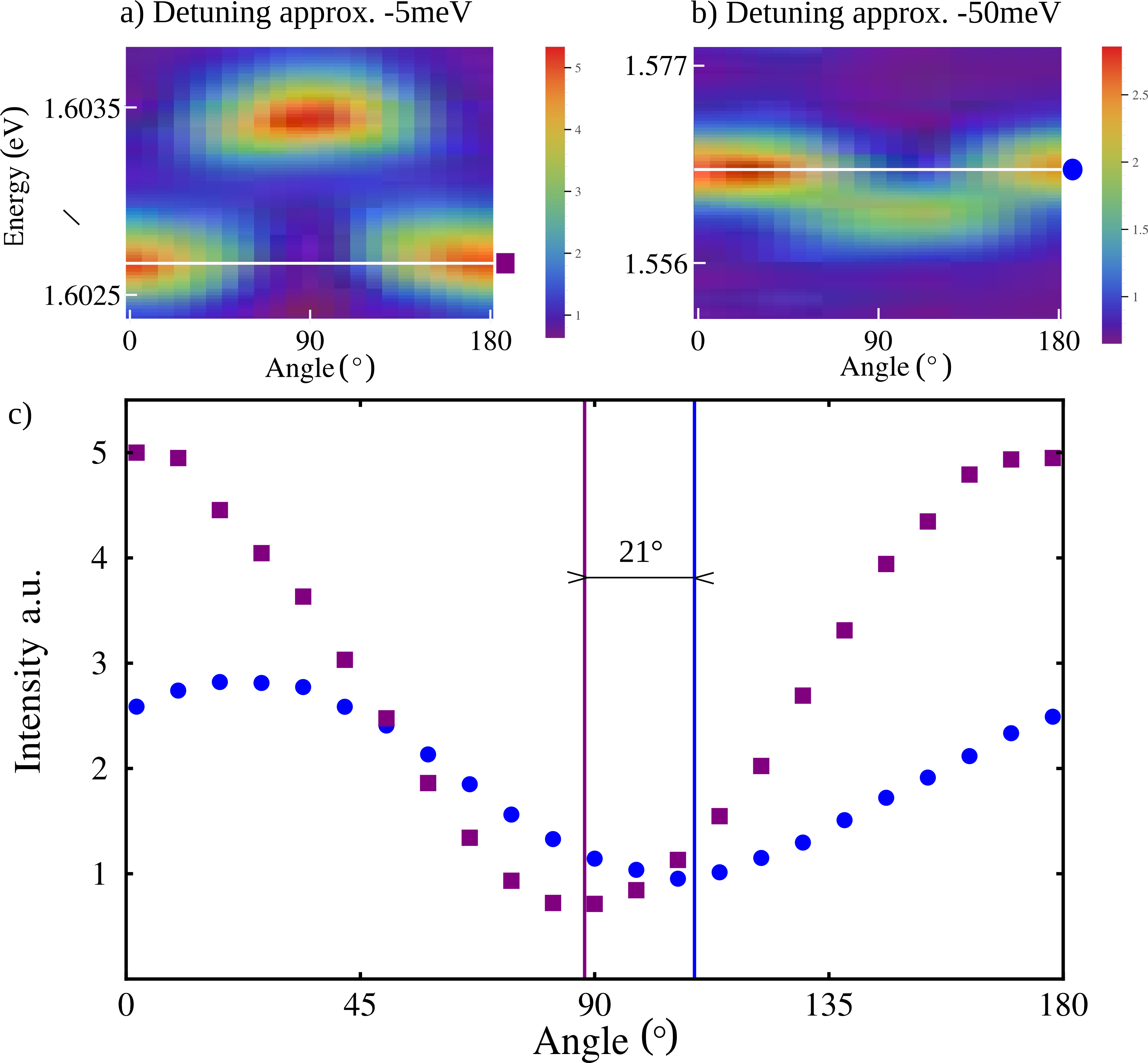}
\protect\caption{\label{fig:angle} PL at normal incidence as a function of polarization angle for (a) small and (b) large negative detuning. $0\degree$ (resp. $90\degree$) polarization angle corresponds to polarization parallel (resp. perpendicular) to the wire long axis. Each column corresponds to the energy spectrum at normal incidence ($\theta=0$). The energy splitting is directly accessible and is for (a) $\delta E_\mathrm{pol}=\unit{0.77}{meV}$ and for (b) $\delta E_\mathrm{pol}=-\unit{0.23}{meV}$. c) Intensity profile as a function of the polarization angle along the white lines in a) (circles) and b) (squares). The polarization basis at large negative detuning (b) is shifted by an angle of \unit{-21}{\degree} with respect to polarization basis parallel and perpendicular to the wire axis. More generally, angular shift varies from \unit{-13}{\degree} to \unit{-21}{\degree} depending on the choice of energy cut and the residuals of the sinusoidal fit are significant for the large negative detuning implying that an unknown systematic effect is probably responsible for the drift and spanning of values compared to the theoretical one which is about \unit{-6}{\degree}.}
\end{figure}
\end{center}

From the model described in appendix~D, we predict a slight variation of the principal polarization
axes with the detuning which is indeed observed experimentally on Fig.~\ref{fig:angle}. Therefore, depending on the detuning, the polarization basis relevant for the splitting is not always perfectly parallel and orthogonal to the wire axis.
This can be explained as follows: since the primitive
cell deformation is along the wire direction only, the principal polarization axes for the exciton eigenstates are parallel and orthogonal
to the wire axis. However, the birefringence induced principal axes
are determined by the wire direction and the relative primitive cell
orientation. Therefore the principal axes related to the photonic
effect are not necessarily exactly parallel and orthogonal to the
wire axis. As $\delta E_{\mathrm{ph}}$, $\delta E_{\mathrm{exc}}$ and $\delta\Omega$
weights on the resulting splitting $\delta E_{\mathrm{pol}}$ depends on the
detuning, so does the principal axis of the polariton branches. 
Note that the polariton polarizations are always linear due to the time-symmetric nature of the system considered. 

In principle the polarization splitting expression~(\ref{eq:splitting}) considered in section~\ref{sec-Contributions} has to be corrected to take into account the various linearly polarization bases, but in practice the precisions of the experimental data is insufficient to infer the introduced extra parameters in the fitting model due to vanishingly small corrections.

\subsection{Polarization splitting due to confinement of the cavity mode}

For lateral modes $j\neq 0$, the polarization splitting increase beyond the maximal splitting of the $j_0$ mode (for example in figure \ref{fig:Splitting-1}). This imply that the lateral confinement also results in an additional polarization splitting. 

Its effect on excitonic levels is extremely weak, below \unit{2}{\micro \electronvolt} for the main contribution due to long range exchange interaction, and can be safely neglected. 
Lateral confinement implies that the TE-TM splitting of the cavity mode comes into play and introduces an extra photonic splitting~\cite{Maragkou2011}, see appendix~E.
For realistic wires, this effect is negligible: below \unit{0.1}{\milli \electronvolt} for the worst case scenario considered in this work ($j_2$ mode of \unit{3}{\micro \meter}-wide wires, $(E_\mathrm{ph} - E_{\mathrm{DBR}})=\unit{-0.1}{\electronvolt}$).

Finally, the TE and TM modes couple differently to the exciton level (see appendix~E) and this results in an additionnal polarization Rabi splitting:

$$ \delta \Omega \sim \frac{\Omega}{2} (n_1^{-2}+2 n_2^{-2}+n_c^{-2}) (\frac{\hbar \pi c (j+1)}{W E_c})^2, $$

where $n_1=n_c$ (respectively $n_2$) is the $\textrm{Al\ensuremath{{}_{0.95}}Ga\ensuremath{{}_{0.05}}As}$ (resp $\textrm{Al\ensuremath{{}_{0.2}}Ga\ensuremath{{}_{0.8}}As}$ ) refractive index, $W$ is the wire width, $E_c$ is the cavity mode energy, and $c$ is the speed of light in the vaccuum. 

This polarization splitting is negligible for $j_0$ lateral modes, but significant for $j>1$ (up to \unit{-0.47}{\milli \electronvolt} for the $j_2$ mode of \unit{3}{\micro \meter}-wide wires).

\begin{center}
\begin{figure}[!h]
\includegraphics[width=8.6cm]{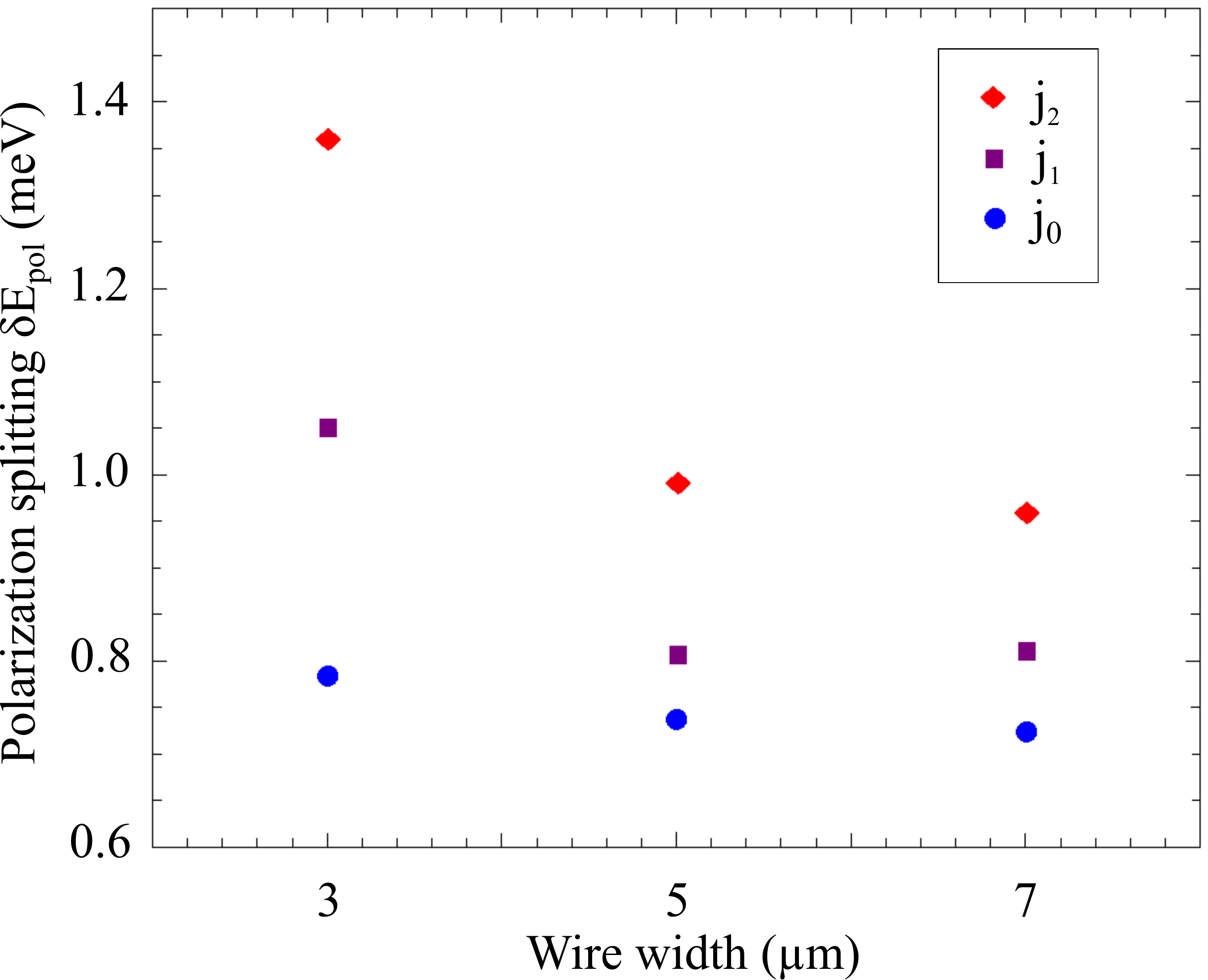}
\protect\caption{\label{fig:LatConf} 
Polarization splitting $-\delta E_{\mathrm{pol}}$ for wire widths $w= 3,5$, and \unit{7}{\micro \meter} and for lateral modes $j_0,\, j_1,\ \mathrm{and}\ j_2$. Measures are done for a reference exciton-photon detuning of the $j_0$ lateral mode about \unit{-13}{\milli \electronvolt}. Note that the effective exciton-photon detuning varies with the lateral mode index for a fixed wire width. (see section~\ref{ssec-Temp})
 }
\end{figure}
\end{center}

In figure \ref{fig:LatConf}, the polarization splitting $\delta E_{\mathrm{pol}}$ is reported 
for the lateral modes $j=0,1, \mathrm{and}\ 2$ and various wire widths.

As anticipated, the polarization splitting is larger for narrow wires and large j. 
The exact  $\frac{(j+1)^2}{w^2}$  dependence is hidden in fig.~\ref{fig:LatConf} due to the competing effect of exciton-photon detuning (which increases with lateral mode j) on polarization splitting. 
The largest polarization splitting observed is \unit{-1.36}{\milli \electronvolt} for $j_2$ and $w=\unit{3}{\micro \meter}$, while the maximum value for $j_0$ with the same wire at null detuning is \unit{-0.95}{\milli \electronvolt}, indicating that the confinement is responsible for at least \unit{-0.41}{\milli \electronvolt} consistently with the anticipated value (\unit{-0.47}{\milli \electronvolt}).

\subsection{Engineering of polaritonic wires}

There are three possibilities to engineer the polarization splitting $\delta E_{\mathrm{pol}}$ of the fundamental mode $j=0$:  
First, by decreasing the QW thickness, the light hole-heavy hole energy splitting $\Delta E$ is increased thus reducing $\delta \Omega$ and $\delta E_{\mathrm{exc}}$. Since the QW thickness also governs the exciton binding energy in a nonlinear fashion, its choice is dictated by the the matching of the excitonic transition with the center of the DBR microcavity stopband. 

Second, by choosing other DBR or wafer composition, the strain $\epsilon_{xx}^{(0)}$ can be changed. This choice is interesting but again, the DBR composition is dictated by the band-gap they provide and engineering the polarization splitting is likely to be a secondary objective. 
Finally, the most promising way is by controlling the depth and direction of etching. As opposed to other control parameters, this can be done once the sample is grown and can be optimized easily on a single wafer. The lateral strain $\epsilon_{xx}^{(0)}$ is solely controlled by the DBR and wafer composition in deep wires, but is lowered for shallow wires due to the nearby influence of the wafer lattice. As explained in appendix~C, the lateral strain is also larger in the [100] direction than in the [110] direction, consequently, a precise choice of polarization splitting can be done by simply tilting slightly the etching direction. Note, however that degeneracy can only be obtained in the [100] or [110] directions (see appendix~D).

Let us finally remark, that the direction dependent-splitting induced by the lattice direction relative to the wire allows to easily design polariton phase gates with S shaped wires or by locally increasing the etching depth along a straight wire in the context of quantum computing with polaritons.~\cite{Solnyshkov2015} Using this technique and taking into account optical spin Hall effect~\cite{Kavokin2005} during polariton propagation, it should be possible to realize any rotation on the Jones sphere.

\section{Conclusion}
\label{sec-Conclusion}

We have measured the microcavity exciton-polariton polarization splitting
in microwires as a function of the temperature, the wires width and the
bare exciton-cavity photon detuning. It appears that the polarization
splitting of the fundamental mode $j_0$ can be controlled and inversed either by controlling the
cavity-exciton detuning or the temperature. 
This observation is explained considering an interplay of two contributions: 
The polarization splitting of the cavity mode and the polarization dependent Rabi splitting. 
The splitting magnitude is controlled by the strain orthogonal to the wire resulting from the lattice mismatch between the AlAs and GaAs layers. It neither depends on the wire width (for narrow wires) nor on external applied constraints on the substrate. Polarization eigenstates non colinear to the wire are observed at large negative detuning and result from the angle between the etching direction and the crystalline axes. Those results are highly reproducible and allow to envision an engineering of the microwires exciton-polaritons eigenstates. 

\begin{acknowledgments}

We are grateful to Robson Ferreira for helpful discussions on our theoretical models. 

\end{acknowledgments}

\appendix

\section{Stress relaxation in the sample}
\label{app:Relaxation}

Previous experimental results on wires presented a polarization splitting interpreted as a consequence of the presence of a stress in the substrate due to thermal contraction of the
sample holder.~\cite{Diederichs2007} As observed in section~\ref{ssec-AppStress}, changing the constraints by sticking differently the sample did not have any influence on the observed splitting.
This apparent contradiction can be explained by a continuum mechanical model. 

The mechanical equilibrium conditions are provided by the Navier equation 
$\vec{0}=\textrm{div}\vec{\vec{\sigma}}+\vec{f}_{v},$
where $\vec{f}_{v}$ are internal volumic forces, zero in our case
and $\vec{\vec{\sigma}}$ is the stress  tensor. The stress tensor
is obtained from the strain tensor $\vec{\vec{\varepsilon}}$ through
Hooke's law: $\sigma_{ij}=C_{ijkl}\varepsilon_{kl},$ 
where $C_{ijkl}$ is the stress-strain tensor and Einstein summation
conventions are used. Al$_{x}$Ga$_{1-x}$As is not an isotropic cristal
but its stress-strain tensor can be approximately considered as such. Hence the particular form for Hooke's law reads 
$\vec{\vec{\sigma}}=\frac{E}{1+\upsilon}(\vec{\vec{\varepsilon}}+\frac{\upsilon}{1-2\nu}\textrm{Tr}(\vec{\vec{\varepsilon}})\vec{\vec{I}}),$
where $E$ is the Young's modulus, $\upsilon$ is the Poisson's ratio, and $\vec{\vec{I}}$ is the identity tensor.

Considering an homogeneous wire we obtain the following mechanical
equations:

\begin{equation}
\begin{cases}
\alpha\partial_{x}^{2}u_{x}+\beta\nabla^{2}u_{x} & =-\alpha\partial_{x}\partial_{z}u_{z}\\
\alpha\partial_{z}^{2}u_{z}+\beta\nabla^{2}u_{z} & =-\alpha\partial_{x}\partial_{z}u_{x}
\end{cases}\label{eq:poisson}
\end{equation}

where $\alpha=E/2(1+\nu)(1-2\nu)$, $\beta=E/2(1+\nu)$ and $\vec{u}$
is the displacement field. The boundary conditions are i)~a vanishing strain
on the free interfaces of the wire and ii)~a linear displacement field
characterized by $u_{x,max}$ at the bottom of the bulk GaAs. 

\begin{figure}[!h]
\includegraphics{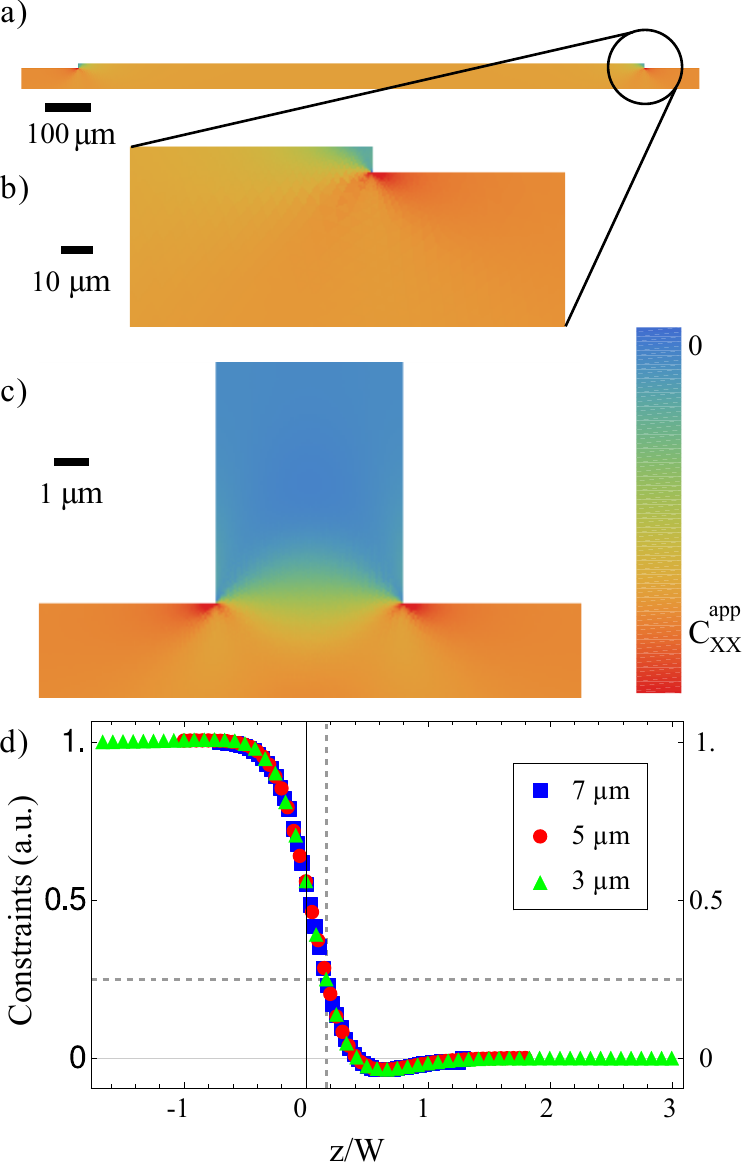}
\protect\caption{\label{fig:contraintes}Constraint repartition in a \unit{7}{\micro\meter}-wide
wire (a). A lateral stress is applied in the bulk (from which
a small part is represented in the bottom part of the sketch) and
is released in the wire. In fig. (b), The lateral stress $C_{XX}$
is plotted against the reduced vertical position $z/W$ where $z$
is the height in the wire and $W$ is the wire width. For narrow wires
($W<H$) the profiles superposition suggests a universal law implying
that the characteristic relaxation length is proportionnal to the
wire width. }
\end{figure}

This linear set of partial differential equations is solved using
a finite element method \footnote{The software used is Cast3M freely available at http://www-cast3m.cea.fr}, the results are presented on fig. \ref{fig:contraintes}. 

For narrow wires, a universal behaviour of the lateral stress relaxation length is observed numerically. 
Although the set of equations is not easily solvable analytically near the connecting edges of the wire due to the singularity of the displacement field, in the wire volume the second member in equation \ref{eq:poisson} can be safely ignored. 
This leads to the following equation which can be used to obtain the typical stress relaxation
length:
$\alpha\partial_{x}^{2}u_{x}+\beta\nabla^{2}u_{x}=0,$
which is a Laplace equation in 2 dimensions. Considering the boundary condition ii) is replaced by a fixed linear displacement field at the interface between the wire and bulk GaAs, the solution of
this equation is known analytically: 
\begin{equation}
u_{x}(x,z)=\frac{-W u_{x,max}}{\pi}\sum_{n=1}^{\infty}\frac{\sin2\pi nx/W\sinh2\pi ncz/W}{n\sinh2\pi ncH/W}\label{eq:3}
\end{equation}
 where $W$ is the wire width, $H$ is the wire length and $c=\sqrt{\alpha/\beta+1}$.

Figure \ref{fig:contraintes} represents a solution for realistic
wire dimensions. 

From eq.~(\ref{eq:3}), we extract a relaxation length
$l=W/2\pi/c=W/2\pi/\sqrt{(2-2\nu)/(1-2\nu)}$. For GaAs $\nu=0.31$
so that $l=W/2\pi/1.91$.

This formula implies that stress in the etching direction is relaxed
on very short lengthscales (typically \unit{0.25}{\micro\meter} for \unit{3}{\micro\meter}
wires and \unit{0.60}{\micro\meter} for \unit{7}{\micro\meter} wires,
the wire height being about \unit{7}{\micro\meter}). This explains
why no difference is observed in our sample between wires of various width and with different sticking methods (section~\ref{ssec-AppStress}).

However, along the wire axis,
the relaxation length is much longer than the wire height and therefore
the bulk stress is preserved through the structure. It is this constraint
anisotropy between the directions parallel and perpendicular to the wire axis which
is the cause of the polarization splitting $\delta E_{\mathrm{pol}}$.

In the sample studied in \cite{Diederichs2007}, the wires were shallower,
they had been etched to slightly below the quantum wells, and the
thermal constraints were effective because they were not relaxed in the lateral direction.

\section{Data analysis}
\label{app:DataAnalysis}

We first briefly recall how raw data analysis is performed (see section~\ref{sec-Origins}) and detail how measurement uncertainties are obtained.

The dispersion relationships similar to the one shown in fig.~\ref{fig:Splitting-1} are experimentally obtained. 
To obtain the $k=0$ lowest polariton branch energy as precisely as possible we make use of the whole polariton branch. Each energy spectrum at fixed polariton momentum $k$ is first fitted using a free Lorentzian with a background varying linearly in energy. The value of the polariton energy for each $k$ along with its uncertainty is obtained from this fitting procedure. The quality of fit also allows to discard lines for which the parasitic light of the pump laser has been too important, this corresponding to 10-20\% of the data points.

The resulting data are fitted using a fourth order polynomial and a weight on data points proportionnal to the above deduced uncertainties. In principle, a more restrictive fitting model using the polariton dispersion branch equation could be used but this model is irrelevant experimentally due to the spherical aberrations of our experimental setup, which play a significant role at large in-plane momentum.

This Lorentzian model matches well experimental data for negative and zero detuning. However, for positive detuning, the polariton branch spectral width becomes extremely large (up to 8 times the polarization splitting) and its curvature nearly vanishes. The fourth order polynomial fitting is therefore not relevant and the error bars given by this method are underestimated. For these detunings, we integrate the 20 lines at $k$ slightly above 0 and fit it by an assymetric lorentzian function, for each of the two polarized branches. The energy values for the minimum of each branch are slightly overestimated since the integration is not done around $k=0$ to avoid the parasitic laser light. Therefore, we use this method to get values and realistic error bars for the energy splitting only. \\

Once the polarization splitting has been carefully extracted from experiments, the fitting procedure can be simplified by considering the Rabi splitting is large compared to the polarization splitting terms (an hypothesis verified \textit{a posteriori}), this leads to eq.~(\ref{eq:splitting}) and (\ref{eq:easy}). The relative uncertainty on $\Omega$ propagates on $\delta \Omega$ due to the induced uncertainty on $Y$. As $\Omega$ is obtained with a typical 2\% standard deviation, its effect on $\delta \Omega$ is negligible compared to (\ref{eq:3}) fitting uncertainties \textit{a posteriori} (about 10\%). This justifies that the fitting procedures for $\Omega$ and $\delta E_{\mathrm{pol}}$ can be done separately.

To estimate the quality of fit, the residuals (not shown) are analyzed : 
(i) Apart from the positive detuning region, residuals are randomely distributed and their autocorrelation function does not reveal any order implying the standard error estimate is not underestimated. (ii) p-values are below 1\% meaning the null hypothesis for fitting parameters is extremely unlikely (note however, that it does not imply that the model fit is correct, see discussion in section~\ref{ssec-Origins-strain}). 
(iii) For positive detunings, residuals are significantly larger meaning that the fitting procedure is not able to capture the details of systematic effects of smaller amplitude, probably secondary anticrossings with other levels (excitonic, traps, etc... ).

This fitting procedure is done using five adjustable parameters ($\Omega$, $E_{\mathrm{exc}}$, $\delta E_{\mathrm{exc}}$, $\delta E_{\mathrm{ph}}$, $\delta\Omega$), and assuming that the detuning calibration has been previously made. By incorporating the detuning calibration in the fitting procedure, the joint fit uses seven adjustable parameters (the previous ones plus the two coefficients of the linear relationship between position on the sample and detuning). Despite this large number, the results are identical to the simple fitting procedure both in estimated values and error bars.

Table~\ref{tab:tabletot} gives the fitting polarization splitting contributions obtained for the various wire widths. In average, the polarization excitonic splitting is $\delta E_{exc}\simeq\unit{-0.54 (20)}{\milli\electronvolt}$, the polarization photonic splitting $\delta E_{ph}\simeq\unit{-0.73 (7)}{\milli\electronvolt}$, and the polarization Rabi splitting is $\delta \Omega\simeq\unit{-1.55 (12)}{\milli\electronvolt}$. All values in Table~\ref{tab:tabletot} are compatible within the error bars, so the wire width indeed has no measurable influence on the various sources of splitting. 

\begin{table}
\begin{tabular}{c|ccccc}
Wire width (\unit{}{\micro\meter}) 				& 3 		& 4 		& 5 		& 6 		& 7 	\tabularnewline
\hline 
\hline 
$\delta E_{\mathrm{exc}}$ 	(\unit{}{\milli\electronvolt})	& -0.56 	& -0.72 	& -0.58 	& -0.49 	& -0.35 	\tabularnewline
$\delta E_{\mathrm{ph}}$	(\unit{}{\milli\electronvolt})	& -0.74		& -0.73 	& -0.76 	& -0.77 	& -0.66 	\tabularnewline
$\delta\Omega$ 			(\unit{}{\milli\electronvolt})	& -1.55 	& -1.61		& -1.62		& -1.52		& -1.46	\tabularnewline
\end{tabular}\protect\caption{Resulting contributions from the different splitting sources obtained by fitting the data for the various wire widths. Typical standard deviations for $\delta E_{\mathrm{exc}}$, $\delta E_{\mathrm{ph}}$, and $\delta\Omega$ are 0.20, 0.07 and \unit{0.12}{\milli\electronvolt} respectively . As observed on fig.~\ref{fig:Stick-1}, estimated parameters are independent of the wire width. \label{tab:tabletot} 
}
\end{table}

\section{Mechanical model of strain distribution in microwires}
\label{app:LatticeMismatch}

The effect of the lattice mismatch can be modeled as follows: 
let us consider the stiffness tensors C for the two kinds of layers of the DBR. 
Due to the cubic symmetry of the lattice cell, they both read $\mathrm{C}_{\mathrm{\bold{11}}} \delta_{ij}\delta_{kl}+\mathrm{C}_{\mathrm{\bold{12}}} \delta_{ijkl}+ \mathrm{C}_{\mathrm{\bold{44}}} (\delta_{ik}\delta_{jl}+\delta_{il}\delta_{kj})(1-\delta_{ij})(1-\delta_{kl})$.
Now we can write the mechanical equilibrium equations: 
The mechanical equilibrium at the edge of the wire perpendicular to its axis reads
$(C_{ijkl}^{(1)}\varepsilon_{kl}^{(1)}h^{(1)}+C_{ijkl}^{(2)}\varepsilon_{kl}^{(2)}h^{(2)})v_{i}^{\bot}v_{j}^{\bot}=0,$ 
where indices (1) and (2) stand for 
$\textrm{Al\ensuremath{{}_{0.2}}Ga\ensuremath{{}_{0.8}}As}$ 
and 
$\textrm{Al\ensuremath{{}_{0.95}}Ga\ensuremath{{}_{0.05}}As}$ DBR layers respectively, $h^{(1,2)}$ are the respective thicknesses of the layers, $\vec{v}^{\bot}$ the unit vector orthogonal to the wire etching direction in the plane of the DBR, and $\varepsilon_{ij}^{(1,2)}$ are the strain tensors already defined in appendix~A.

The mechanical equilibrium of the (1)-(2) interface in the vertical direction gives the condition: 
$(C_{zzkl}^{(1)}\varepsilon_{kl}^{(1)}+C_{zzkl}^{(2)}\varepsilon_{kl}^{(2)})=0$. 
The relative lattice mismatch $\Lambda^{(2)}-\Lambda^{(1)}$ between layers (2) and (1) at the interfaces gives the continuity condition: 
$\varepsilon_{ij}^{(2)}=\varepsilon_{ij}^{(1)}+(\Lambda^{(1)}-\Lambda^{(2)}) \delta_{ij},$
for i,j in the heterostructure plane. 

Since the lattice parameter of (1) (material of the bulk) is fixed in the wire direction we have
$\varepsilon_{ij}^{(1)}v_{i}^{\Vert}= -\Lambda^{(1)} \delta_{ij} v_{i}^{\Vert}$ for j in the plane. 

The GaAs quantum well strain tensor $\varepsilon_{ij}^{(0)}$ has a lattice parameter fixed in the wire direction: 
$\varepsilon_{ij}^{(0)}v_{i}^{\Vert}= 0$ for j in the plane. 

The quantum well (0) is also in mechanical equilibrium with (2), the intracavity material, so that 
$(C_{zzkl}^{(0)}\varepsilon_{kl}^{(0)}+C_{zzkl}^{(2)}\varepsilon_{kl}^{(2)})=0$

Finally, the invariance of strain along y and x direction leads to 
$\varepsilon^{(n)}_{xz}=\varepsilon^{(n)}_{yz}=0$. 

All those latter equations lead to strain tensors of the form: 
$$\varepsilon_{ij}^{(n)}=
\begin{pmatrix}
-\Lambda^{(n)} + \varepsilon \cos ^2 \theta & -\varepsilon \cos \theta \sin \theta& 0\\
-\varepsilon \cos \theta \sin \theta & -\Lambda^{(n)} + \varepsilon \sin ^2 \theta & 0\\
0 & 0 & \varepsilon^{(n)}_{zz}
\end{pmatrix}$$

where $\varepsilon$ is the resulting strain orthogonal to the wire axis (which is identical in all layers), 
$\theta$ is the angle between the etching direction and the [100] crystalline direction of the bulk GaAs, 
$\Lambda^{(n)}$ is the relative change in lattice parameter between bulk GaAs and the considered AlGaAs ternary alloy layer (which are 0, 270 and 1300 ppm for layers 0,1 and 2 respectively),
 and $\varepsilon_{zz}^{(n)}$ are known functions of $\varepsilon,\Lambda^{(n)},h$ and $\theta$. 
The resulting orthogonal strain $\varepsilon$ is obtained by minimizing the free energy $F$ of the whole crystal considering that it consists mainly in DBR pairs: 
$F=C_{ijkl}^{(1)}\varepsilon_{ij}^{(1)}\varepsilon_{kl}^{(1)}h^{(1)}/2+C_{ijkl}^{(2)}\varepsilon_{ij}^{(2)}\varepsilon_{kl}^{(2)}h^{(2)}/2$.

The resulting expression can be easily computed but has a tedious expression and we  will just comment its main properties. 
First, the orthogonal strain $\varepsilon$ is proportionnal to the relative change in lattice parameters $\Lambda^{(n)}$. 
Second, the orthogonal strain depends on the angle $\theta$ between the wire etching direction and the crystal lattice in the following way : 
$$\varepsilon^{-1} (\theta) =  \frac{\varepsilon^{-1}  (0) +\varepsilon^{-1}  (\frac{\pi}{4}) }{2}+\frac{\varepsilon^{-1}  (0) -\varepsilon^{-1}  (\frac{\pi}{4}) }{2} \cos 4\theta$$
with $\varepsilon^{-1}  (0) < \varepsilon^{-1}  (\frac{\pi}{4})$ (due to the stiffness tensor parameter values). 
This means that the orthogonal strain is larger when the wire is etched along the [100] or [010] direction than when it is etched along a diagonal [110] direction.

Using the values referenced in \cite{Adachi1985} for strain-stress tensor components and lattice parameters at room temperature, and a wire direction of $\theta \simeq \unit{39}{\degree}$ with respect to the crystalline axis (measured by optical microscopy), a resulting strain $\varepsilon \simeq 900$ ppm is obtained.  

\section{Derivation of the polarization splittings induced by a lateral strain.}
\label{app:theory}

\subsection{Polarization photonic splitting}

The photonic eigenmode energies are determined by the refractive index of the DBR layers. 
The total induced birefringence on the cavity mode (neglecting the intracavity layer thickness and assuming that the electric field is evenly distributed on the 2 kinds of layers of the DBR) is given by 
$\delta\eta_{ij} / \epsilon^2=\frac{1}{h^{(1)}+h^{(2)}} (h^{(1)} U_{ijkl}^{(1)} \varepsilon_{mn}^{(1)} +
 h^{(2)} U_{ijkl}^{(2)} \varepsilon_{mn}^{(2)}),$ 
where $\epsilon$ is the average DBR permittivity,
and $U_{ijkl}^{(1)}$ is the photoelastic tensor of material (1) which has the same symmetry as the stiffness tensor $C_{klmn}^{(1)}$ described in the previous section (idem for layer (2)). 

First, let us remark that matrices $\varepsilon_{ij}^{(1,2)}$ as well as the photoelastic tensor are symmetric. Consequently, the induced birefringence tensor is symmetric. Moreover, one can easily show that the $z$-axis is an eigenvector of the induced birefringence. Consequently, the in-plane eigenvectors of the induced birefringence are orthogonal and the polarization basis of the induced photonic splitting is linear. 

Second, if the photoelastic tensor components are such that $\mathrm{U}_{\mathrm{\bold{12}}} \neq 2\mathrm{U}_{\mathrm{\bold{44}}}$ (which is in general the case), the polarization basis for the photonic state is not colinear to the wire etching direction, accordingly to experimental observations (see fig.~\ref{fig:angle}). 
The angle of deviation $\xi$ at large photonic detuning between the observed polarization basis and the wire etching basis is then dependent on the angle $\theta$.
Note that no deviation should be observed for [100], [010] and [110] directions.

The  polarization splitting amplitude can be theoretically computed, using the formula 
$$ \delta E_{\mathrm{ph}} = \frac{E_{\mathrm{ph}} \delta \eta}{2 \bar n},$$
where $\bar n$ is effective refractive index in the cavity mode and $\delta \eta$ is the maximal average variation in electric susceptibility induced by strain in material (1) and (2) which is simply the difference in the eigenvalues of $\delta\eta_{ij}$ restricted to the microcavity plane. 
Estimates of the photoelastic coefficients based on experiments at room temperature \cite{Adachi1985} lead to a typical $\delta E_{\mathrm{ph}}$ in the $\unit{-0.2}{\milli\electronvolt}$ region,  i.e. of the same order of magnitude than than the  \unit{-0.73 (7)}{\milli\electronvolt} photonic polarization splitting experimentally observed. A precise determination is not possible due to both the lack of measurements of the photoelastic coefficients at low temperatures and the extreme sensitivity of those coefficients to the bandgap energy. 

\subsection{Excitonic eigenstates}

The Hamiltonian for the exciton states is given by

$$H_0+H_{\mathrm{exch}}+H_{\mathrm{PB}}$$

Here $H_0$ is the exciton Hamiltonian which is diagonal in the heavy exciton - light exciton basis and associates an energy $E_{hh}$ ($E_{lh}$) to heavy hole-excitons (light hole-excitons respectively). It takes the form $H_0 = -\frac{1}{2}\Delta E (J_z^2-J^2/3)$, where $\Delta E = E_{lh} - E_{hh}>0$, and J are angular momentum operators acting on the hole states. It results from two contributions: (i)~The difference in confining energy of holes in the QWs.  The confining potentiel is \unit{474}{\milli\electronvolt} for a \unit{7}{\nano\meter} GaAs QW with $\textrm{Al\ensuremath{{}_{0.95}}Ga\ensuremath{{}_{0.05}}As}$ barriers. By solving the finite potential well problem~\cite{Fishman2010} using standard Luttinger parameters~\cite{Madelung2004}, we obtain a light-hole heavy-hole difference in confinement energy of \unit{29.3}{\milli\electronvolt}. (ii)~The binding energy of heavy and light hole excitons is slightly different due to their different Bohr radii \unit{9.25}{\nano \meter} and \unit{9.52}{\nano \meter} respectively (see appendix~F). This results in an additional \unit{-2.0}{\milli\electronvolt} contribution.~\cite{Bastard1992} The total heavy exciton - light exciton splitting is thus $\Delta E =${27.3}{\milli\electronvolt}.

$H_{\mathrm{exch}}$ is the short range exchange interaction between the electron and hole of the exciton. 
Defining the uncoupled hole-electron basis,  $| \frac 3 2, \up \ket$, $| \frac 3 2, \down \ket$, $| \frac 1 2, \up \ket$, $| \frac 1 2, \down \ket$, $| -\frac 1 2, \up \ket$, $| -\frac 1 2, \down \ket$, $| -\frac 3 2, \up \ket$, $| -\frac 3 2, \down \ket$, where the first index is the magnetic number of the hole state and the second one is the electron spin state in the $\hat z$ basis. 
In this basis $H_{\mathrm{exch}}$ reads~\cite{Chen1988, Maialle1993}

$$\frac{3}{4} \frac{\Delta E_{\mathrm{SR}}}{|\phi^{\mathrm{3D}}(0)|^2}
\begin{pmatrix}
0 & 0 & 0 & 0 & 0 & 0 & 0 & 0\\
0 & I_{hh} & -\frac{I_{hl}}{\sqrt{3}} & 0 & 0 & 0 & 0 & 0\\
0 & -\frac{I_{hl}}{\sqrt{3}} & \frac{I_{ll}}{3} & 0 & 0 & 0 & 0 & 0\\
0 & 0 & 0 & \frac{2 I_{ll}}{3} & \frac{-2 I_{ll}}{3} & 0 & 0 & 0\\
0 & 0 & 0 & \frac{-2 I_{ll}}{3} & \frac{2 I_{ll}}{3} & 0 & 0 & 0\\
0 & 0 & 0 & 0 & 0 & \frac{I_{ll}}{3} & -\frac{I_{hl}}{\sqrt{3}} & 0\\
0 & 0 & 0 & 0 & 0 & -\frac{I_{hl}}{\sqrt{3}} & I_{hh} & 0\\
0 & 0 & 0 & 0 & 0 & 0 & 0 & 0
\end{pmatrix}
$$

where $\Delta E_{\mathrm{SR}}$ is the short range exchange splitting in the bulk semiconductor \cite{Maialle1993}, 
$\phi^{\mathrm{3D}}(0)$ is the 3D hydrogenic exciton wave function at zero relative distance,
$I_{jj'}$ is a form factor defined in appendix~F.

Finally, $H_{\mathrm{PB}}$ is the Pikus-Bir deformation Hamiltonian given by 

\begin{eqnarray*}
H_{\mathrm{PB}} = a_v (\epsilon_{xx} + \epsilon_{yy} + \epsilon_{zz}) &+ b [ (J_x^2-J^2/3) \epsilon_{xx}+ c.p. ]+ \\
 &\frac{2 d}{\sqrt{3}}  [ \frac{1}{2} (J_x J_y + J_y J_x) \epsilon_{xy} + c.p. ],
\end{eqnarray*}

where $a_v$, $b$ and $d$ are deformation potentials, $\epsilon_{ij}$ are the strain tensor components of the GaAs quantum well material (0) obtained in appendix~C, and c.p. designate cyclic permutations of $x,y$ and $z$.

Before diagonalizing this Hamiltonian, we first note that only states with a total magnetic number $\pm 1$ are optically active due to optical selection rules. 
Moreover, bright and dark excitons are not mixed by the exchange interaction or the Pikus Bir deformation Hamiltonian since shear is absent (see appendix~C). 
We can consequently work on the bright exciton subspace in which the total Hamiltonian takes the form:

$$\frac{1}{2}
\begin{pmatrix}
-A & -B & C & 0\\
-B & A & 0 & C\\
C & 0 & A & -B\\
0 & C & -B & -A
\end{pmatrix}
$$

in the basis $| \frac 3 2, \down \ket$, $| \frac 1 2, \up \ket$, $| -\frac 1 2, \down \ket$, $| -\frac 3 2, \up \ket$, 
where 
\begin{eqnarray*}
A &=&\Delta E + b \epsilon + (\frac{I_{ll}}{3}-I_{hh}) \frac{3}{4} \frac{\Delta E_{\mathrm{SR}}}{|\phi^{\mathrm{3D}}(0)|^2} \\
B &=&\frac{2}{\sqrt{3}}I_{hl} \frac{3}{4} \frac{\Delta E_{\mathrm{SR}}}{|\phi^{\mathrm{3D}}(0)|^2}\\
C &=&\sqrt{3} b \epsilon .
\end{eqnarray*}

The symmetry of this Hamiltonian allows to reduce again the dimensionality of the problem by going in the linearly polarized basis $| H, X \ket$, $| L, X \ket$, $| H, Y \ket$, $| L, Y \ket$. 
In this basis, the Hamiltonian can be written

$$\frac{1}{2}
\begin{pmatrix}
-A & C-B & 0 & 0\\
C-B & A & 0 & 0\\
0 & 0 & -A & -B-C\\
0 & 0 & -B-C & A
\end{pmatrix}.
$$

 Linearly polarized states are defined as 
\begin{eqnarray*}
|H,X/Y\ket =  |- \frac 3 2, \up\ket \pm | \frac 3 2, \down \ket \\
|L,X/Y\ket =  |\frac 1 2, \up \ket \pm | -\frac 1 2, \downarrow \ket ,
\end{eqnarray*}

The eigenstates are two linearly polarized excitons mainly heavy-hole ($|H,X\ket$ and $|H,Y\ket$) and two linearly polarized excitons mainly light-hole ($|L,X\ket$ and $|L,Y\ket$).
We are interested in the mainly heavy hole excitons which form the polaritons. Their splitting $\delta E_{\mathrm{exc}} \simeq \frac{BC}{A}$ reaches \unit{-0.38}{\milli \electronvolt}, compatible with the experimentally infered value of \unit{-0.54 (20)}{\milli \electronvolt}.

The unnormalized eigenstates at the lowest order in $(B+C)^2/A^2$ read

\begin{equation}
|E_B,X/Y\ket = 2 A |H,X/Y\ket - (C \mp B ) |L,X/Y\ket 
\label{eq:eigenstates}
\end{equation}

The electric dipole transition for each state is proportionnal to the matrix element 
$|\bra \Psi^e|\vec{\epsilon}\cdot \vec{p}|\Psi^h\ket|$, where $|\Psi^{e/h}\ket$ are the electron and hole wave-functions, $\vec{\epsilon}$ is the polarization of the light field and $\vec{p}$ is the momentum operator. 
The Rabi splitting for both polarizations reads~\cite{Fishman2010}

\begin{eqnarray*}
\Omega_X &=& K (2A D_{hh} + (C-B) D_{lh})\\
\Omega_Y &=& K (2A D_{hh} - (B+C) D_{lh}),
\end{eqnarray*}
where $K$ is a common positive constant incorporating the enveloppe wave-function contribution and the cavity characteristics, and $D_{hh/lh}$ are the norms of the p-matrix elements of the Bloch-wave functions. 
The obtained polarization splitting is 
$$\delta \Omega \simeq  2 \Omega C  \frac{1}{2\frac{D_{hh}}{D_{lh}}A-B},$$
where the heavy-hole light-hole exciton dipolar transition ratio $D_{hh}/D_{lh} \sim  \sqrt{3}$ is mainly determined by the Clebsch-Gordan coefficients for the heavy and light-hole composition (while form factors only result in a few percent correction).

As can be observed from this result, the polarization Rabi splitting does not involve exchange interaction. 
As observed in \cite{Balili2010} it results from the heavy hole - light hole mixing due to the effect of uniaxial stress, but as can be seen from eigenstates~\ref{eq:eigenstates}, the light exciton fraction is almost equal in both states and the resulting polarization splitting results from the interference between the light exciton and heavy exciton dipoles.

For the \unit{5}{\micro\meter}-wide wires $\Omega=$\unit{16.1 (3)}{\milli\electronvolt}, so that $\delta \Omega$ is expected to reach \unit{-1.0}{\milli\electronvolt} (using the complete theoretical expression) close from the infered value $\delta \Omega=$\unit{-1.62 (12)}{\milli\electronvolt}.

\section{Effects of confinement on the Rabi polarization splitting}
\label{app:theoryconf}

For $j>0$, additional contributions to the polarization splitting due to confinement have to be taken into account. First, let us define here the effective confinement angle $\theta_c$ : 
The lateral confinement implies that the cavity modes are plane waves in the lateral direction indexed by the lateral mode number $j=0,1,...$. 
Neglecting the contribution of the evanescent field outside of the wire, the lateral wavevector reads 

$$k_x = \frac{\pi (j+1)}{W}$$

Since the vertical wavevector $k_z$ is $\frac{n_c E_c}{\hbar c}$, where $n_c$ is the cavity layer refractive index (labeled $n^{(2)}$ in appendix~C), $E_c$ is the cavity mode energy, and $c$ is the speed of light in the vacuum, and $W$ is the wire width. 
We can then define the effective confinement angle as

$$\sin \theta_c = \frac{k_x}{k_z} = \frac{\hbar c \pi (j+1)}{W n_c E_c}.$$

which is the equivalent angle to the normal direction of the virtual propagating plane-wave corresponding to the laterally confined mode considered. 


This effective angle implies that TE-TM splitting of the cavity mode is responsible for an extra photonic splitting given by~\cite{Panzarini1999} 

\begin{equation}
\frac{2 L_c L_{\mathrm{DBR}}(0)}{(L_c + L_{\mathrm{DBR}}(0))^2} (E_\mathrm{ph} - E_{\mathrm{DBR}}) \sin^2 \theta_c
\end{equation}

where $L_c$ is the length of the Fabry-P\'erot cavity, 
$L_{\mathrm{DBR}}$ is the penetration depth of the field in the DBR, 
and $ E_{\mathrm{DBR}}$ is the center of the stopband energy of the DBR. 


Lateral confinement affects the coupling between the cavity modes, TE ($\parallel$) and TM ($\perp$), and the excitonic mode~\cite{Panzarini1999} which reads
\begin{eqnarray*}
\Omega^{\TE} (\theta) &=& \Omega \sqrt \frac{L_{\mathrm{eff}}(0)}{L_{\mathrm{eff}}^{\TE}(\theta)} \frac{1}{\cos \theta_c}\\
 \Omega^{\TM}  (\theta) &=& \Omega \sqrt \frac{L_{\mathrm{eff}}(0)}{L_{\mathrm{eff}}^{\TM}(\theta)},
\end{eqnarray*}
where $\theta$ is the angle of the incoming field outside of the cavity, $\theta_c = \sin^{-1} (\frac{1}{n_c} \sin \theta)$ is the equivalent angle in the cavity layer, $L_{\mathrm{eff}}^{\TE/\TM}(\theta)$ is the effective cavity length $L_c+ L_{\mathrm{DBR}}^{\TE/\TM}(\theta)$. The penetration $L_{\mathrm{DBR}}^{\TE/\TM}(\theta)$ of the TE/TM mode in the DBrs reads

\begin{eqnarray*}
L_{\mathrm{DBR}}^{\TE}(\theta) &=&
\frac{2 n_1^2 n_2^2 (a+b)}{n_c^2 (n_2^2-n_1^2)}
\frac{\cos^2 \theta_1 \cos^2\theta_2}{\cos^2\theta_c}
\\
L_{\mathrm{DBR}}^{\TM}(\theta) &=&
\frac{2 n_1^2 n_2^2 }{n_c^2}
\frac{a \cos^2 \theta_1 + b \cos^2\theta_2}{n_2^2 \cos^2\theta_1-n_1^2 \cos^2\theta_2},
\end{eqnarray*}

where $\theta_{1} = \sin^{-1} (\frac{n_0}{n_1} \sin \theta)$ (respectively $\theta_{2}$) are the equivalent angles to the input angle $\theta$ after refraction in the layer of index $n_1$ ($n_2$), 
and $a$ ($b$) is the thickness of the layer 1 (2) of the DBR. 
Note that in these formula $n_1<n_2$, whereas in appendix~C, exponents label the layer sequence so that for the cavity considered in this work $n_1 =n_c \equiv n^{(2)}$, $n_2 \equiv n^{(1)}$, $a \equiv h^{(2)}$, and $b \equiv h^{(1)}$.

After some algebra, the equivalent behavior for small angles is obtained: 

\begin{eqnarray*}
\frac{L_{\mathrm{DBR}}^{\TE}(\theta)}{L_{\mathrm{DBR}}(0)}
 &\sim&
1-(n_1^{-2}+n_2^{-2}-n_c^{-2}) n_c^2 \sin^2 \theta_{c}
\\
\frac{L_{\mathrm{DBR}}^{\TM}(\theta)}{L_{\mathrm{DBR}}(0)} &\sim&
1+(1-(\frac{n_1}{n_2})^2)(\frac{b}{a+b}+\frac{n_1^2}{n_2^2-n_1^2}) \sin^2 \theta_1
\end{eqnarray*}

For AlGaAs microcavities, in $L_{\mathrm{DBR}}^{\TM}$, the last term prevail over the geometric term $b/(a+b)$ so that

$$\frac{L_{\mathrm{DBR}}^{\TM}(\theta)}{L_{\mathrm{DBR}}(0)} \sim
1+\frac{n_c^2}{n_2^2} \sin^2 \theta_c
$$

Introducing these approximate expressions in the coupling term and considering $L_{\mathrm{DBR}}$ is much larger than $L_c$ we obtain the polarization Rabi splitting: 

$$ \delta \Omega \sim \Omega (n_1^{-2}+2 n_2^{-2}+n_c^{-2}) \frac{n_c^2}{2}\sin^2 \theta_{\mathrm{c}}.$$

\section{Theoretical parameters determination}
\label{app:values}

In this appendix the various parameters values used in our theoretical expressions are detailed.
The deformation potentials $a_v$, $b$ and $d$ are \unit{1.16}{\electronvolt}, \unit{-1.7}{\electronvolt}, and \unit{-4.55}{\electronvolt} respectively. \cite{Chuang1995a}

The computation of the form factors implied in the short range exchange interaction require the determination of the envelope wave-functions. 
We used values tabulated in \cite{Madelung2004} for conduction electron, heavy, and light hole effective masses. 
The confining potential for holes and electrons is \unit{474}{\milli\electronvolt} and \unit{711}{\milli\electronvolt} according to Bastard \cite{Bastard1992}.

With these values, the confinement energy for heavy and light holes is found to be \unit{17}{\milli\electronvolt} and \unit{45.7}{\milli\electronvolt} respectively by solving the finite potential problem. \cite{Fishman2010} 
The binding energies for heavy and light hole excitons are found using charts in \cite{Bastard1992} and are \unit{8.9}{\milli\electronvolt} and \unit{10.}{\milli\electronvolt} respectively using an effective well thickness to account for the evanescent excitonic field outside of the well. 

The form factor $I_{j'j}$ (where $j$ and $j'$ indicate the light or heavy hole character of the exciton) is given by the expression~\cite{Maialle1993}: 
$$
I_{j'j} = \phi_{1s(j')} (0) \phi_{1s(j)} (0)
\int \mathrm{d} z |\xi_{1c} (z)|^2 \xi_{1j'h} (z) \xi_{1jh} (z),
$$
where $\phi_{1s}$ are the 2D envelope wave-functions of the considered excitons, 
and $\xi_{1c}$ ($\xi_{1jh}$) is the conduction envelope wavefunction along the growth axis in the conduction (respectively heavy, or light hole) subband. 
Using the aforementioned wavefunctions, $I_{hl}$ values is found to be $\unit{1.18 \, 10^{-3}}{\nano \meter \rpcubed}$.
$I_{ll}$ and $I_{hh}$ have similar values $\unit{1.12 \, 10^{-3}}{\nano \meter \rpcubed}$ and $\unit{1.26 \, 10^{-3}}{\nano \meter \rpcubed}$. 

$1/|\phi^{\mathrm{3D}}(0)|^2 = \pi a_0^{*3} \simeq \unit{9518}{\nano\meter \cubed }$, where $a_0^*$ is the 3D Bohr radius of the GaAs exciton. 

The bulk short range exchange energy $\Delta E_{SR}$ used in this work is \unit{0.37}{\milli \electronvolt} according to \cite{Gilleo1968}. Note that experimental determination of the short range exchange energy is difficult and experimental values span from \unit{0.02}{\milli \electronvolt} according to Eckardt et al. \cite{Ekardt1979} to \unit{0.37}{\milli \electronvolt} according to Gilleo et al. \cite{Gilleo1968}. (see \cite{Chen1988} for a more detailed discussion) Using the relations and experiments in this work, we can deduce an estimate of the bulk short range exchange energy of \unit{0.50(20)}{\milli \electronvolt}.

Using the above values, the effective light hole - heavy hole exciton splitting $A$ (see appendix~D) is found to be \unit{25.0}{\milli \electronvolt}.

Finally refractive indexes at low temperatures of 3.008 for $\textrm{Al\ensuremath{{}_{0.95}}Ga\ensuremath{{}_{0.05}}As}$ and 3.458 for $\textrm{Al\ensuremath{{}_{0.2}}Ga\ensuremath{{}_{0.8}}As}$ have been used.

\end{document}